\documentclass[conference]{IEEEtran}

\makeatletter

\usepackage{algorithm}
\usepackage{algpseudocode}
\usepackage{blindtext}
\usepackage{subfig}
\usepackage{eso-pic}
\usepackage{fancyhdr}
\IEEEoverridecommandlockouts
\usepackage{cite}
\usepackage{amsmath,amssymb,amsfonts}
\usepackage{graphicx}
\usepackage{textcomp}
\usepackage{xcolor}
\def\BibTeX{{\rm B\kern-.05em{\sc i\kern-.025em b}\kern-.08em
    T\kern-.1667em\lower.7ex\hbox{E}\kern-.125emX}}
    
\usepackage{eso-pic}

\newcommand{\Aspace}{\mathcal{A}}
\newcommand{\Sspace}{\mathcal{S}}

\def\1{\mbox{1\hspace{-0.25em}l}}
    
 \makeatletter
\def\ps@IEEEtitlepagestyle{%
  \def\@oddhead{}%
  \def\@evenhead{}%
  \def\@oddfoot{\hfill\thepage\hfill}%
  \def\@evenfoot{\hfill\thepage\hfill}%
}
\makeatother   
\begin{document}
\title{\vspace*{1cm} Optimal Execution with Reinforcement Learning in a Multi-Agent Market Simulator
}

\author{\IEEEauthorblockN{Yadh Hafsi}
\IEEEauthorblockA{\textit{CMAP} \\
\textit{École Polytechnique}\\
Paris, France \\
https://orcid.org/0009-0001-0686-0349}
\and
\IEEEauthorblockN{Edoardo Vittori}
\IEEEauthorblockA{
\textit{Intesa Sanpaolo}\\
Milan, Italy \\
https://orcid.org/0000-0003-4648-1797}
}

\maketitle

\begin{abstract}
This study investigates the development of an optimal execution strategy through reinforcement learning, aiming to determine the most effective approach for traders to buy and sell inventory within a finite time horizon. 
Our proposed model leverages input features derived from the current state of the limit order book and operates at a high frequency to maximize control. To simulate this environment and overcome the limitations associated with relying on historical data, we utilize the multi-agent market simulator ABIDES, which provides a diverse range of depth levels within the limit order book. We present a custom MDP formulation followed by the results of our methodology and benchmark the performance against standard execution strategies. Results show that the reinforcement learning agent outperforms standard strategies and offers a practical foundation for real-world trading applications.
\end{abstract}



\section{Introduction}
\label{sec:intro}

Research in market microstructure shows that large trades influence asset prices because the immediate depth of the market is limited (see \cite{BOUCHAUD2009}); a single large order can exhaust all current buyers or sellers. 
This suggests that it is generally advantageous to split large orders into several smaller blocks.

There is an extensive stream of literature on how to optimally split orders, which will be introduced in Section \ref{sec:related}, but the majority of this research restricts market behavior to Brownian motion with hard coded market impact assumptions and does not simulate the liquidity present in the markets, obtaining an execution trajectory which is a function of time. 

Reinforcement learning (RL) is a very flexible machine learning paradigm which can be applied to learn data-driven execution policies, without having to make specific assumptions on market behavior. 
On the other hand, RL algorithms are very data hungry and require thousands of simulations to be trained, for this reason having a market simulator is essential.

One can simulate the market using stochastic processes and making assumptions on the market impact, but then the optimal approach is to use the traditional solutions which are optimal in that context \cite{macri2024reinforcement}. Another solution can be to use historical data, as is done by several in the optimal execution with RL framework (\cite{hendricks2014reinforcement, ning2018double}), which gives access to LOB information, but in general enables only a simulation of temporary impact as the price must revert back to the historical dataset while in a realistic framework there is also temporary and transient impact (see Section \ref{sec:impact}).

The approach considered in this paper is through the use of a multi agent market simulation, which enables the simulation of the entire LOB and takes into account the impact of the trades in the market.
Examples of such simulators for LOB modeling include \cite{Alfi2008MinimalAB2,chakraborti2011econophysics,hamill2015agent, Coletta2022LearningTS}. This development has facilitated multi-agent approaches like \cite{balch2019evaluate} and led to the creation of ABIDES \cite{byrd2019abides} which is the simulator considered in this work.

In this paper we present a framework based on RL, specifically the Double Q Network (DQN) \cite{ddqn} algorithm, to learn an optimal execution strategy. Training and testing occur on synthetic data generated through the multi-agent market simulator ABIDES. 

The rest of the paper is organized as follows, Section \ref{sec:related} reviews the related works, Section \ref{sec:optim_exec_intro} introduces the limit order book, market impact, the multi agent simulation and then the optimal execution problem setting. 
Section \ref{sec:RL_definitions} introduces RL and explains how the optimal execution problem can be solved with this technology.
Section \ref{sec:experimental_results} presents our experimental results.

\section{Related Works} \label{sec:related}

We will divide the literature on optimal execution into two major strands: the classical mathematical finance framework and the more recent RL approaches.
Within the mathematical finance tradition, \cite{bertsimas1998optimal} first established a model where prices follow Brownian motion with linear permanent and temporary market impact, deriving the Time-Weighted Average Price (TWAP) as the optimal solution. 
This was extended by \cite{almgren2001optimal}, who augmented the objective function with the minimization of the variance of the expected shortfall, yielding an analytical solution that incorporates risk aversion—resulting in more front-loaded execution profiles. 
\cite{obizhaeva2005optimal} further refined the model by introducing transient impact that decays over time, producing a U-shaped execution trajectory with higher trading intensity at the beginning and end of the period.
Building on this, Busetti and Lillo \cite{busseti2012calibration} introduced risk aversion into the transient impact model, confirming the U-shaped profile but with heavier emphasis on early execution. 
\cite{cartea2015algorithmic} expanded this line of work by testing alternative objective functions, showing that inventory penalties act analogously to risk aversion and exploring terminal penalties for unfinished execution.

A common feature across these studies is their focus on what we may term \textit{strategic execution}: the optimal trading trajectory depends solely on deterministic variables such as time, permanent impact, and risk aversion, without regard to real-time market microstructure or limit order book dynamics. The execution period lasts a few hours, with usually five minute time-steps.
The other type of execution which can be analyzed is the \textit{tactical execution}, where the focus is to optimize the trade decisions within each period. In this case, LOB information is fundamental.

Reinforcement learning methods allow for such tactical execution strategies, thus we will now analyze them by dividing into the two categories. 
The strategic RL approaches have in general the objective of proving RL can find the optimal solutions found by traditional methods. 
In this category are the following papers: \cite{hendricks2014reinforcement, ning2018double, macri2024reinforcement}, which typically operate over long time windows and use algorithms such as Q-learning, DQN, or PPO to learn execution policies from historical or simulated limit order book data. 

Conversely, tactical RL approaches deal with shorter time horizons and more granular decision-making. Early work by Nevmyvaka et al. \cite{nevmyvaka2006reinforcement} pioneered this direction, optimizing execution over five-minute intervals. Subsequent contributions \cite{pardo2022modular, moallemi2022reinforcement, Nagy2023}, investigated higher frequency control problems such as deciding between market and limit orders, stopping execution under uncertainty, or leveraging predictive signals to guide trades.

In this work, we aim to unify these two traditionally separate layers of the execution problem. By training a reinforcement learning agent to operate over a 30-minute window with one-second control intervals, we enable the simultaneous learning of both strategic scheduling and tactical decision-making within a single framework, bridging the gap between high-level trade planning and low-level order placement in real time.

The most similar to our approach is \cite{karpe2020multi}, as they use ABIDES as the market simulator and they have a high (30 second) control frequency to learn both a strategic and tactical execution. 
Nonetheless, their experimental results are unsatisfactory as they state that their RL agent solution converges to the TWAP.
Instead, in this work we have run an extensive experimental campaign obtaining results superior to the TWAP and other baseline strategies.

\section{Optimal Execution Setting}\label{sec:optim_exec_intro}
\subsection{Limit Order Book}\label{sec:LOB_intro}

\begin{figure}[ht]
\centering
\includegraphics[width=0.75\linewidth]{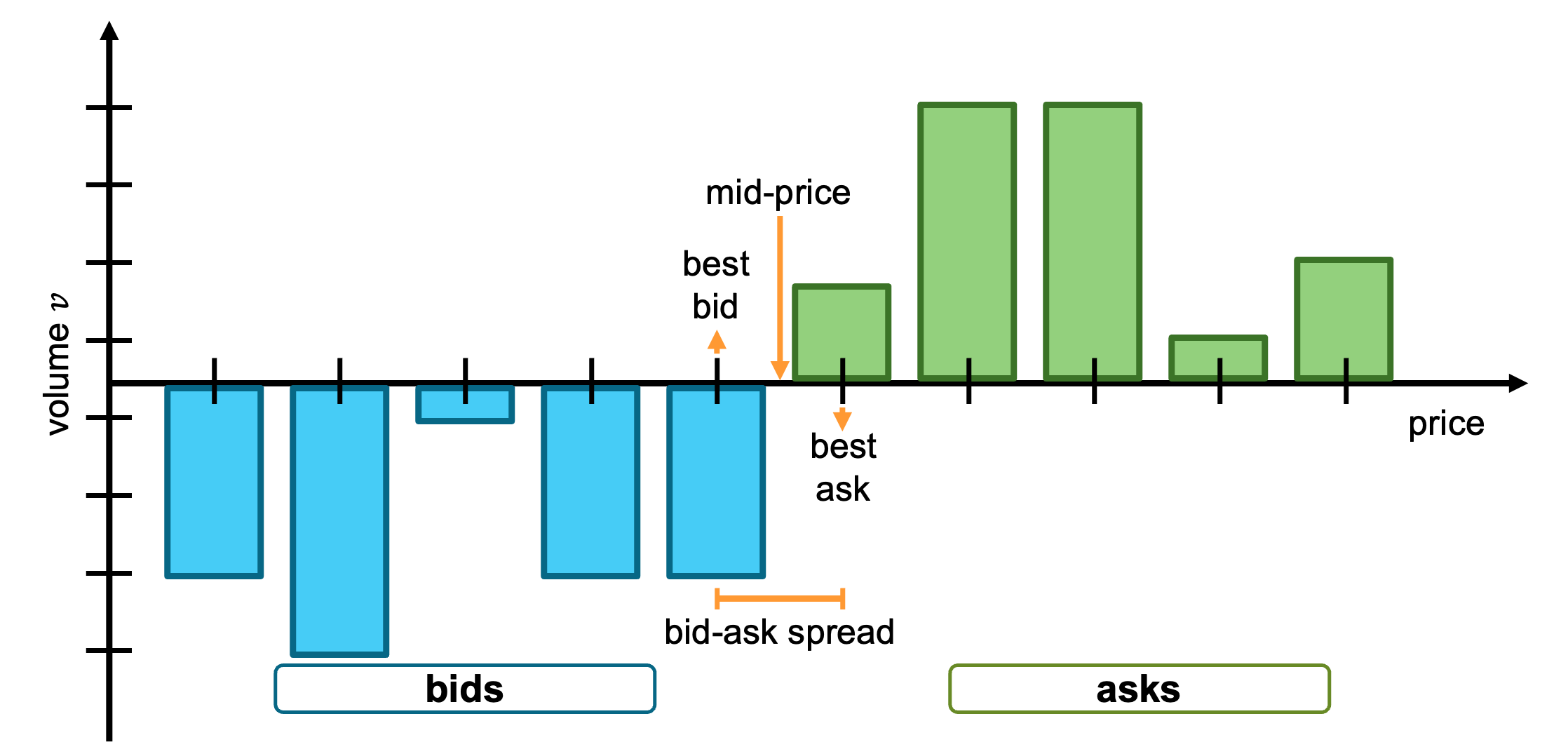}
\caption{Representation of the Limit Order Book}
\label{fig:LOB}
\end{figure}

A limit order book is a comprehensive record of current limit orders, where price changes occur discretely in increments known as the \textit{tick size} (see Figure \ref{fig:LOB}). 
Each order consists of a price and size, collectively contributing to the volume represented by $Q$. 
One of the key elements highlighted in the LOB is the concept of \textit{liquidity}. 

\noindent
In a perfectly liquid market, any quantity of a specific security can be immediately converted to cash. Measurement of liquidity involves metrics such as traded value/turnover, bid-ask spread, and trade-through intensity, as described in \cite{chevalier2023uncovering}. 
\paragraph{Order Types} \label{sec:orders}
When trading in a LOB, it is possible to employ different types of orders.
The limit order is the most used and entails specifying the price at which we want to execute the trade. 
There is no guarantee that the trade will happen, as the limit price may never be reached. 
These orders are executed considering a price-time priority, also known as First-In-First-Out (FIFO).
Then there is the market order, which consists of a request to carry out the order immediately at the best price available in the market. To be more specific, a buy (sell) order is matched with limit sell (buy) orders starting with the best ask price. Finally, there are other types of orders, such as fill-or-kill orders and others.

\paragraph{LOB features} \label{sec:LOB_features}
In algorithmic trading, it is popular to use signals derived from specific characteristics, including the following:
\begin{itemize}
    \item \textit{Total depth}: it corresponds to the cumulative sum of volume at multiple price levels $d$ of the LOB:
    \begin{align*}
        TD_{{h}}^k = & \sum_{j=1}^k Q_{{h}}^j \; \text{for} \; k\in\{1,...,d\}, \; {h} \in \{\text{bid, ask}\},
    \end{align*}
     where $Q_{\text{ask}}^k$ ($Q_{\text{bid}}^k$) is the volume of outstanding limit orders at the $k$-th best ask (bid) price level.
    \item \textit{Volume imbalance}: it describes the difference between the existing order volume on the bid and ask price levels. It can be defined as:
    \begin{equation*}
        v_{h}^{k} = \frac{{TD}^k_{h}}{{TD}^k_{\text{ask}}+{TD}^k_{\text{bid}}} \; \text{for} \; k\in\{1,...,d\}, \; {h} \in \{\text{bid, ask}\}.
    \end{equation*}
    \item \textit{Mid price}: it is the midpoint between the best bid and best ask prices:
    \begin{equation*} \label{eq:mid_price}
        P_{\text{mid}} = \frac{P_{\text{best ask}}+P_{\text{best bid}}}{2}.
    \end{equation*}
    \item \textit{Spread}: it is the difference between the best ask and the best bid:
        \begin{equation*} \label{eq:spread}
        P_{\text{mid}} = P_{\text{best ask}}-P_{\text{best bid}}.
    \end{equation*}
\end{itemize}

\subsection{Market impact} \label{sec:impact}
Market impact refers to the effect that trades have on asset prices and is a key consideration in the design of optimal execution strategies. The literature typically distinguishes between three primary forms of market impact:

\begin{itemize}
    \item \textbf{Temporary Impact:} This refers to the immediate but short-lived distortion in price caused by a trade. It is generally attributed to the consumption of available liquidity at the best bid or ask and tends to vanish quickly as new orders replenish the book or the market reverts. Temporary impact is typically modeled as a cost that affects only the execution price of the current trade.

    \item \textbf{Permanent Impact:} Permanent impact captures the long-lasting effect that a trade has on the asset price. It is often interpreted as the market incorporating the information revealed by the trade, such as the presence of an informed trader or a shift in supply and demand. Permanent impact is usually modeled as a linear function of cumulative volume and affects the price trajectory permanently.

    \item \textbf{Transient Impact:} Transient impact lies between the two extremes above. It reflects a price impact that decays gradually over time rather than disappearing instantaneously (as with temporary impact) or persisting indefinitely (as with permanent impact). This type of impact is useful for modeling situations where market participants adjust their behavior in response to past trades, leading to delayed and decaying price effects. Transient impact is often captured through decaying kernels or memory functions in execution models.
\end{itemize}
\begin{figure}[H]
    \centering
        \includegraphics[scale=0.25]{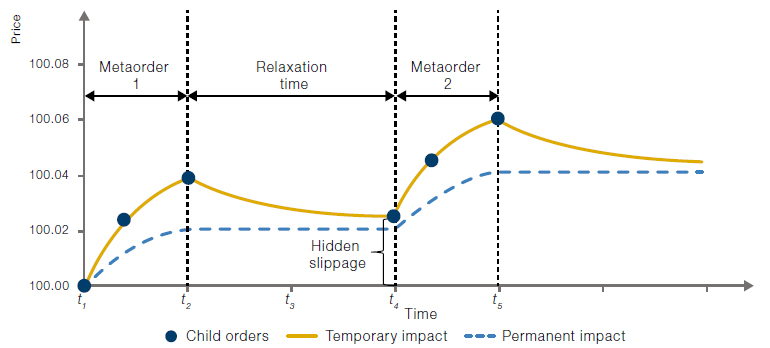}
       \caption{\centering Stylized representation of average permanent vs. temporary price impact (see \cite{harvey2021}).}
\end{figure}
ABIDES reproduces realistic market impact because price responses emerge endogenously from the interaction of heterogeneous agents rather than from imposed functional forms. 
Large orders deplete liquidity at the best quotes, prompting market makers and other agents to adjust their behavior. 
This naturally generates short-lived distortions resembling temporary impact as well as longer-lasting adjustments consistent with permanent and transient impact. 
Thus, the dynamics of impact arise from microstructural interactions, providing a realistic environment for testing execution strategies.
\subsection{Configuration of Multi-Agent Simulations}
Let $(\Omega ,\mathbb{F}, \mathcal{F} = \{\mathcal{F}_t\}_{t \geq 0},\mathbb{P})$ be a complete filtered probability space endowed with right-continuous filtration $\{\mathcal{F}_t\}_{t \geq 0}$. We assume that the filtration $\{\mathcal{F}_t\}_{t \geq 0}$ supports a standard $\mathbb{P}$-Brownian motion $W$ and a random Poisson process $N$ with intensity $\lambda$. In the simulation, various agents, such as Exchange Agents, Adaptive Agents, Market Maker Agents, Value Agents, and Momentum Agents, develop their trading strategies based on an estimate of the fundamental value, which is defined by an Ornstein-Uhlenbeck (OU) process. The dynamics of the fundamental process process can be described by the following stochastic differential equation (SDE):
\begin{equation*}
    \mathrm{d} X_t = \theta (\mu - X_t) \, \mathrm{d} t + \sigma \, \mathrm{d} W_t + J \, \mathrm{d} N_t,
\end{equation*}
where:
\begin{itemize}
    \item $\theta$ is the rate of mean reversion,
    \item $\mu$ is the long-term mean,
    \item $\sigma$ is the volatility,
    \item $J$ is the jump size, drawn from a bimodal distribution centered at zero. Specifically, $J$ can take values from two Gaussian distributions: $\mathcal{N}(\mu_1, \sigma_1^2)$ and $\mathcal{N}(\mu_2, \sigma_2^2)$, where $\mu_1 = -\mu_2$ and $\sigma_1 = \sigma_2$, ensuring the overall mean of the bimodal distribution is zero.
\end{itemize}
The jump process $N$ is introduced to represent the impact of the arrival of major news events on the fundamental value of the traded asset. This allows us to obtain improvements of the simulation computation time while producing more realistic price time series.

We chose the RMSC-4 configuration, which is the reference configuration in ABIDES and was also used in \cite{ABIDESgym} and \cite{zijian}.
Several key settings are defined within the RMSC-4 configuration to realistically simulate market behaviors and participant interactions. Exchange agents manage order books and historical data streams, maintaining a depth of 10 levels and a history of 500 streams. 

Noise agents introduce randomness into the market through the actions of 1000 agents. 

Value agents, numbering 102, base their trades on a true mean fundamental value, which behaves like an Ornstein-Uhlenbeck (OU) process centered around \( \mu_{va} \) of \$100,000. This process has a mean reversion parameter \( \theta_{va} \) set at \( 1.67 \times 10^{-15} \) and an arrival rate \( \lambda_{va} \) of \( 5.7 \times 10^{-12} \).

An oracle-like agent tracks the mean-reversion process with a parameter \( \theta_{or} \) set at \( 1.67 \times 10^{-16} \) and a volatility of \( 5 \times 10^{-10} \) for the fundamental time series, providing a benchmark for the fundamental value.

Market Maker agents dynamically adjust their pricing strategies. They use an adaptive window size, set their order size to $0.025$ percent of the volume, maintain a price range within 10 ticks, and operate with a wake-up frequency of $1$ second to update their quotes. This setup uses a higher frequency compared to other standard configurations, making it more realistic and better adapted to today's markets (e.g. \cite{Nagy2023,karpe2020multi}).

Momentum agents, totaling $12$, trade according to recent market trends, reacting to short-term price movements.

These settings are designed to provide a comprehensive simulation of market dynamics, reflecting the diverse strategies and behaviors of market participants within a controlled environment, as well as simulate realistic market impact.

\begin{figure}[ht]
\centering
\includegraphics[width=0.75\linewidth]{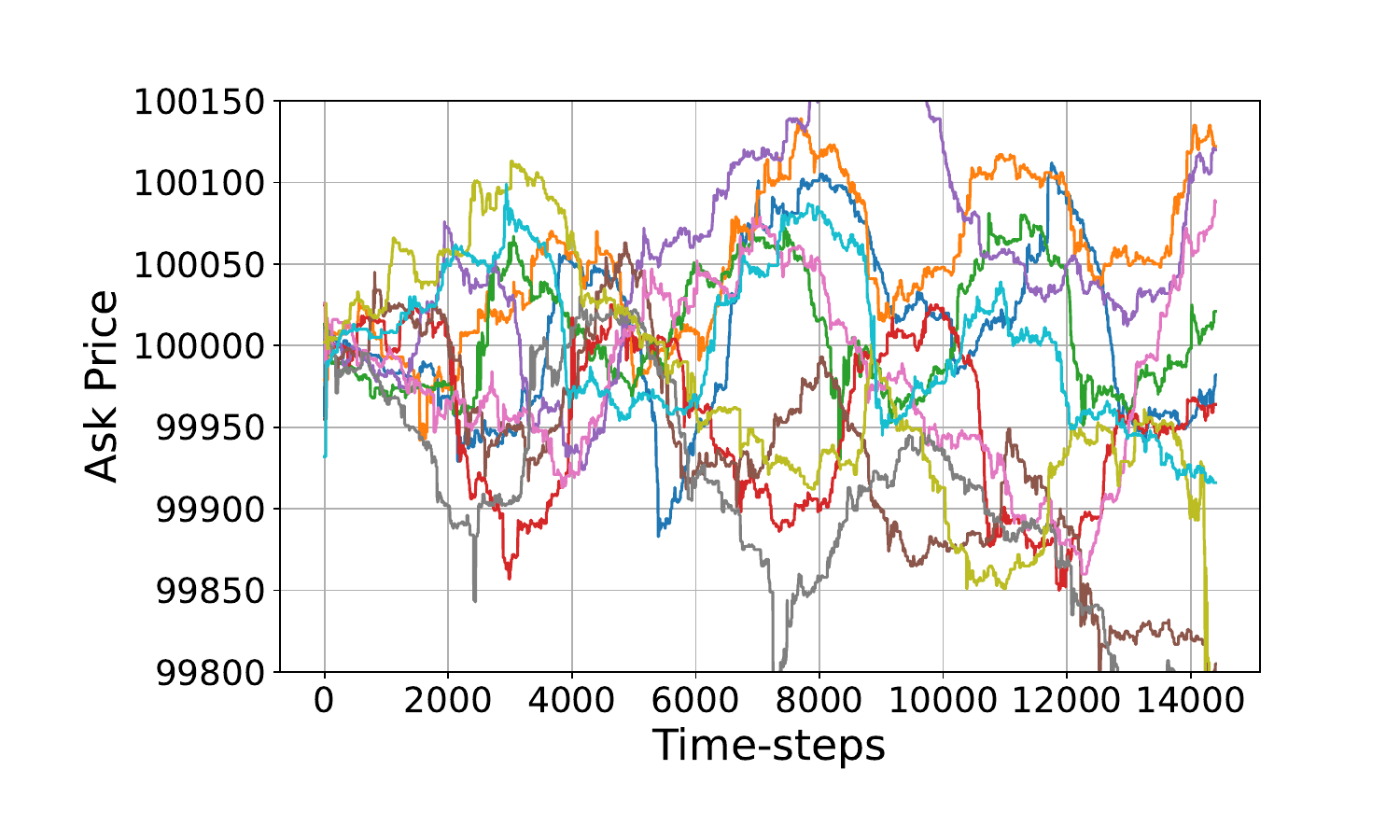}  
\caption{Sample paths of ask prices generated by ABIDES for different seeds.}
\label{fig:ask_price_noexecution}
\end{figure}
\subsection{Problem Formulation}
The optimal execution problem consists in executing a trade of $X_0$ shares over a time interval $[0,T]$. In assessing the trade, it's essential to consider its immediate transactional impact, temporary price fluctuations, and potential long-term effects on future prices due to lasting alterations in pricing dynamics. Additionally, a penalty could be considered if the agent fails to accomplish complete execution on time. 

In a discrete-time setting with $N{+}1$ decision points, the purchasing strategy (analogously for selling) is formulated as a sequential decision process, in which the trader determines the quantity $x_k$ to buy at each time step $t_k = kT/N$, for $k \in \{0, \ldots, N\}$, with $t_0 = 0$ and $t_N = T$. The cumulative purchases form a trajectory $\{x_0, \ldots, x_N\}$, where $x_k$ denotes the cumulative number of shares acquired by time $t_k$. By construction, $x_0 = 0$, and full completion requires $x_N = X_0$ at the terminal time.
Let $P_k$ denote the average execution price for trade $\Delta x_{n} = x_k - x_{k-1}$. The goal is to minimize the expected total execution cost for a risk-neutral trader over a finite time horizon $T$~:
\begin{gather}
\label{control_problem}
\min _{x \in \mathcal{A}} \mathbb{E}\left[\sum_{k=0}^N P_k \Delta x_k\right],\\
\mathcal{A}=\left\{\left\{x_0, x_1, \ldots, x_N\right\}\in \mathbb{R}^{N+1}_+ ; \sum_{k=0}^N | \Delta x_k| =X_0\right\}.
\end{gather}
This cost functional corresponds to the Implementation Shortfall (IS) introduced by \cite{Perold1988}, which measures the excess cost of execution relative to an instantaneous purchase at the initial market price $P_0$. Formally, for a purchasing strategy, the IS is 
\begin{equation*}
\text{IS} = \sum_{k=0}^N P_k \, \Delta x_k - X_0 P_0.
\end{equation*}
Initial studies on optimal execution are based on a stochastic framework in which the dynamics of the assets and execution costs are modeled through SDEs. The first formal analysis of the optimal execution problem is by \cite{bertsimas1998optimal}. Under suitable conditions and using dynamic programming, they provide a closed-form formula as a solution to the optimal execution problem.
Subsequently, an extension of the stochastic model was provided by introducing permanent and temporary effects due to the impacts of orders on the market and by inserting a risk-aversion parameter by \cite{almgren2001optimal}. The Almgren-Chriss model optimizes the trading strategy $x(t)$ over a time horizon $[0, T]$ to minimize total costs, and has a parameter to adjust the trade-off between minimizing market impact and execution costs. 

The most commonly used execution algorithm is the \textit{Time Weighted Average Price (TWAP)}. TWAP aims to execute a trade of size $X_0$ evenly over a specified time horizon $T$ in sizes of $X_0/N$. The arithmetic average of prices collected yields the TWAP price:
\begin{equation} \label{eq:twap}
    \text{TWAP}=\frac{X_0}{N}\sum_{k=0}^N P_{k},
\end{equation}
where $P_k$ is the average at which each trade was executed through a market order sent at time $t_k$. In Almgren-Chriss terms \cite{almgren2001optimal}, TWAP corresponds to a scenario with zero risk aversion, where the trader is indifferent to price volatility and only aims to minimize market impact by trading evenly.

\section{Reinforcement Learning}\label{sec:RL_definitions}
A discrete-time Markov Decision Process (MDP) (see \cite{puterman1990markov}) is defined as a tuple $\langle\Sspace,\Aspace, \mathbb{P}, \mathcal{R}, \gamma, \mu\rangle$, where $\Sspace$ is the state space, $\Aspace$ the action space, $\mathbb{P}(\cdot|s,a)$ is a Markovian transition model that assigns to each state-action pair $(s,a)$ the probability of reaching the next state $s'$, $\mathcal{R}(s,a)$ is a bounded reward function, $\gamma\in[0,1[$ is the discount factor, and $\mu$ is the distribution of the initial state. 
The policy of an agent is characterized by $\pi(\cdot|s)$, which defines for each state $s$ an action with a probability distribution over the action space.
\subsection{Embedding Optimal Execution as an MDP} \label{sec:embedding}

The basic building block for applying RL algorithms is a description of the environment as a MDP:
\begin{itemize}
    \item \textbf{State space:} Percentage holdings remaining, percentage time remaining, volume imbalance up to 5 levels of the LOB, best bid price, best ask price.
    \item \textbf{Action space:} 5 possibile actions: do nothing, consume $Q_t^k = Q_{min}\times k$, $k = 1, \dots, 4$.
    \item \textbf{Reward function:} \begin{equation}
    \label{eq:reward}
    r_{t} =  \underbrace{Q_t^k\times (P_0 - P_t)}_{\text{implementation shortfall}} - \underbrace{\alpha d_t}_{\text{penalty}} - \underbrace{\beta I_T\1_{\{t=T\}}}_{\text{terminal penalty}},
\end{equation}
where $P_t$ is the average execution price, $P_0$ the arrival price, $d_t$ is the depth consumed by the market order at time $t$, and $\alpha>0$, $I_T$ is the remaining inventory at the end of the execution period and $\beta>0$. Once the execution is completed, the reward remains zero for the rest of the period, whereas any residual inventory at time $T$ incurs a terminal penalty.
\end{itemize}
The reward comprises two components: the \textit{implementation shortfall} and a penalty associated with the depth consumed by the executed orders. This configuration allows the agent to develop optimal strategies for acquiring inventory while reducing significant market impact.
Note that we are considering an execution problem in which we want to buy, for this reason we consider $P_0 - P_t$ in the implementation shortfall. 
This is because if we buy at a lower price compared to the arrival price $P_0$, then we lower our execution costs.
If we were considering a sell problem, we would need to invert to $P_t - P_0$.

\subsection{Tailoring the Environment}

Although we have the standard objective in Equation~\eqref{control_problem}, how to appropriately define the MDP to achieve that objective is not straightforward. 
In fact, there are several factors which make that objective challenging:
\begin{enumerate}
\item The asset's price movements are usually independent of your action. 
If/when the action influences the movement of the price, then it is adverse to your objective as you are generating a market impact.
\item It is mandatory to finish the execution by the end of the execution period, but this requirement is hard to embed in the standard reward formulation.
\item If you have a long execution period available, compared to the total execution size, then you can choose whether to execute rapidly, thus increasing market impact, or execute at a slower rate but risking an adverse market movement.
\end{enumerate}
To tackle the first factor, we included the penalty in the reward function, which has the objective of emphasizing that we want to avoid market impact.
To address the second, we include an additional large penalty if the execution is not finished by the end of the total available period. 
To address the third factor, the penalty in the reward function is enabling.
In fact, thanks to the penalty, the reward is negative in most cases.
As a 0 reward obtained consistently once the execution is finished is preferable to a negative reward, the learning agent should be incentivized to finish earlier. 

We also explored different state and action space setups, and those described in Section~\ref{sec:embedding} gave the best results. 
\subsection{Algorithm}
We consider finite horizon problems in which future rewards are exponentially discounted with~$\gamma$.
Let us define a trajectory as a sequence of states, actions, and rewards, up to a stopping time $\tau$:
\begin{equation*}
(s_{0}, a_{0}, r_{1}, s_{1}, a_{1}, r_{2}, ..., s_{\tau-1}, a_{\tau-1}, r_{\tau}).
\end{equation*}

The objective in RL is the maximization of the expected return, given an initial state distribution
\begin{equation*}
    J_\pi := \underset{s_0\sim \mu}{\mathbb{E}_{\pi}}\Big[\sum_{i=1}^{\tau} \gamma^{i-1} r_{i}.\Big],
\end{equation*}
where the return is the discounted sum of the rewards. Numerous RL algorithms are available (see \cite{PPO,schulman2015trust,mnih2015human}). In this work, we focus on the Deep Q-Network (DQN) algorithm, a model-free, online, off-policy reinforcement learning approach, as it provided the best learning results. The DQN algorithm is detailed in algorithm~\ref{algo:DQN}.
\begin{algorithm}
\caption{Deep Q-Network}
\begin{algorithmic}[1]
\State Initialize replay memory $D$ to capacity $N$ 
\State Initialize action-value function $Q$ with random weights
\For{episode = 1 to M}
    \State Initialize state $s_1$
    \For{t = 1 to T}
        \State With probability $ \epsilon$ select a random action $a_t$
        \State Otherwise select $a_t = \arg\max_a Q(s_t, a; \theta)$
        \State Execute action $a_t$
        \State Observe reward $r_t$ and next state $s_{t+1}$
        \State Store transition ($s_t, a_t, r_t, s_{t+1})$ in $D$
        \State Sample random minibatch of transitions $(s_j, a_j, r_j, s_{j+1})$ from $D$
        \State Set $y_j = r_j + \gamma \max_{a'} Q(s_{j+1}, a'; \theta)$
        \State Perform a gradient descent step on $(y_j - Q(s_j, a_j; \theta))^2$
    \EndFor
\EndFor
\end{algorithmic}
\label{algo:DQN}
\end{algorithm}

\section{Experimental Results} \label{sec:experimental_results}

We defined the environment to be as realistic as possible, by using a 1 second control frequency, so that the agent has the maximum flexibility in deciding when to execute. Furthermore, we gave a long maximum execution period so as to give it the possibility to choose how much to make the execution last. 
In the results that follow, we focus on a buy execution, but the results hold also for a sell execution where the only thing to be changed is the sign of the implementation shortfall in the reward.


\subsection{Environment setup}
The environment includes several parameters and variables. The total size of the order to execute is fixed at $20000$ shares. 
The direction specifies whether the parent order is a buy or sell, defaulted to buy. 
A \textit{timeWindow} of $30$ minutes is allotted for the agent to execute the entire order, with a time step duration set to $1$ second. 
We set the incremental size $Q_{min} = 20$ for the buy or sell orders placed by the agent. 
Finally, the penalty is a constant amount imposed per non-executed share at the end of the time window, set at a default of $5$ per share. Additionally, a penalty of $5$ per share is also applied for over-execution beyond the intended order size. The $\alpha$ for the depth penalty in the reward is set to $2$.

\subsection{RL algorithm setup}
We ran a hyperparameter tuning which gave the following optimal parameters.
We employed DQN, with a neural network architecture that features fully connected $2$ input layers composed of $50$ and $20$ neurons.
The learning rate schedule implements a linear decrease from $10^{-3}$ to $0$ in $90,000$ steps, while the exploration rate follows a $\epsilon$-Greedy search from $1$ to $0.02$ in $10,000$ steps. 
In addition, we implement a state history length of $4$ and a market data buffer of length $50$. 
This approach is aimed at ensuring the stability of DQN.
Finally, the objective function is a discounted sum of rewards with a discount factor of $\gamma = 0.9999$. This means that the agent treats all future rewards almost as equally important which is useful in execution problems where rewards come at the end (e.g., penalizing slippage at end of episode).

We also experimented with alternative state spaces (continuous and discrete) and algorithms such as TRPO, PPO, and actor–critic, but DQN consistently proved the most reliable.

\subsection{Baseline algorithms}
We utilize the following baseline algorithms to benchmark the performance of the RL policy:
\begin{itemize}
    \item Time-Weighted Average Price (TWAP): defined in equation \eqref{eq:twap}, corresponding to the risk-neutral solution of the Almgren--Chriss model.
    \item Passive Policy (PP): this policy mostly keeps the agent inactive, with a 60\% chance of doing nothing. With a 40\% chance, it randomly chooses and executes a quantity among four available options, each with equal likelihood.
    \item Random Policy (RP): sample whether to use a market order or do nothing (quantity of 0). There is a 50\% chance of doing nothing. Alternatively, it randomly selects and executes one of three actions with equal probability: do nothing, or consume $Q_t^k = Q_{min}\times k$, with $k = 1, \dots, 3$.
\end{itemize}
    
\subsection{Experiments}
Next, we evaluate the RL agent's ability to execute large orders while minimizing market impact. This involves a detailed comparison of the RL agent's performance with the baseline strategies mentioned earlier.
\begin{figure}[ht]
\centering
\includegraphics[width=0.55\linewidth]{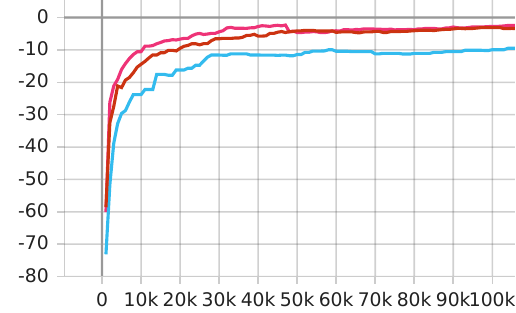}
\caption{Average episode reward for DQN agent with a environment seed equal to 10 and learning rates equal to $10^{-2}$ (pink), $10^{-3}$ (red), and $10^{-4}$ (blue).}
\label{fig:learning}
\end{figure}

\subsubsection{Standard Configuration Analysis}
Figure \ref{fig:learning} shows the learning curve of the DQN algorithm.
The increasing reward curve indicates that the environment is correctly formulated and that the agent is maximizing implementation shortfall and reducing market impact during training, approaching a zero average reward as it learns.



\begin{figure}[ht]
\centering
\includegraphics[width=0.45\linewidth]{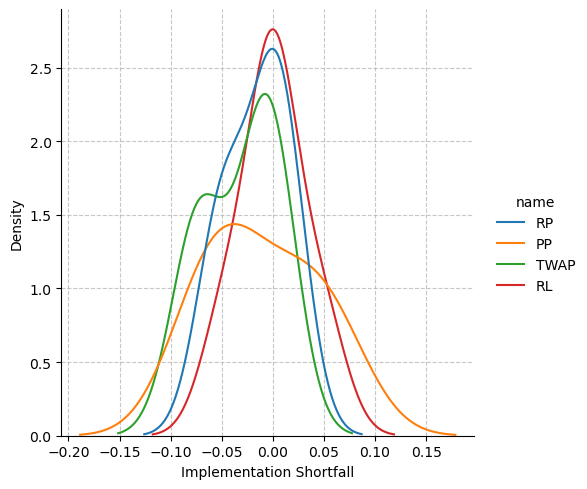}  
\caption{Implementation shortfall normalized by the initial order size distribution.}
\label{fig:pnl_reward}
\end{figure}
In Figure \ref{fig:pnl_reward}, we evaluate out-of-sample the optimized reinforcement learning policy and compare it to the three baselines. The distribution is created after running 20, $30$-minute windows with different seeds. We observe the average step implementation shortfall (see Equation \eqref{eq:reward}).
In this case we see that the RL agent generates a lower variance compared to the baselines with also a higher average. 
This suggests that the RL execution strategy is capable of executing consistently close to the arrival price while minimizing market impact.

\begin{figure*}[!t]
\centering
    \subfloat[Actions taken by the RL agent for a single trajectory over time.\label{fig:prices_actions}]{\includegraphics[scale=0.23]{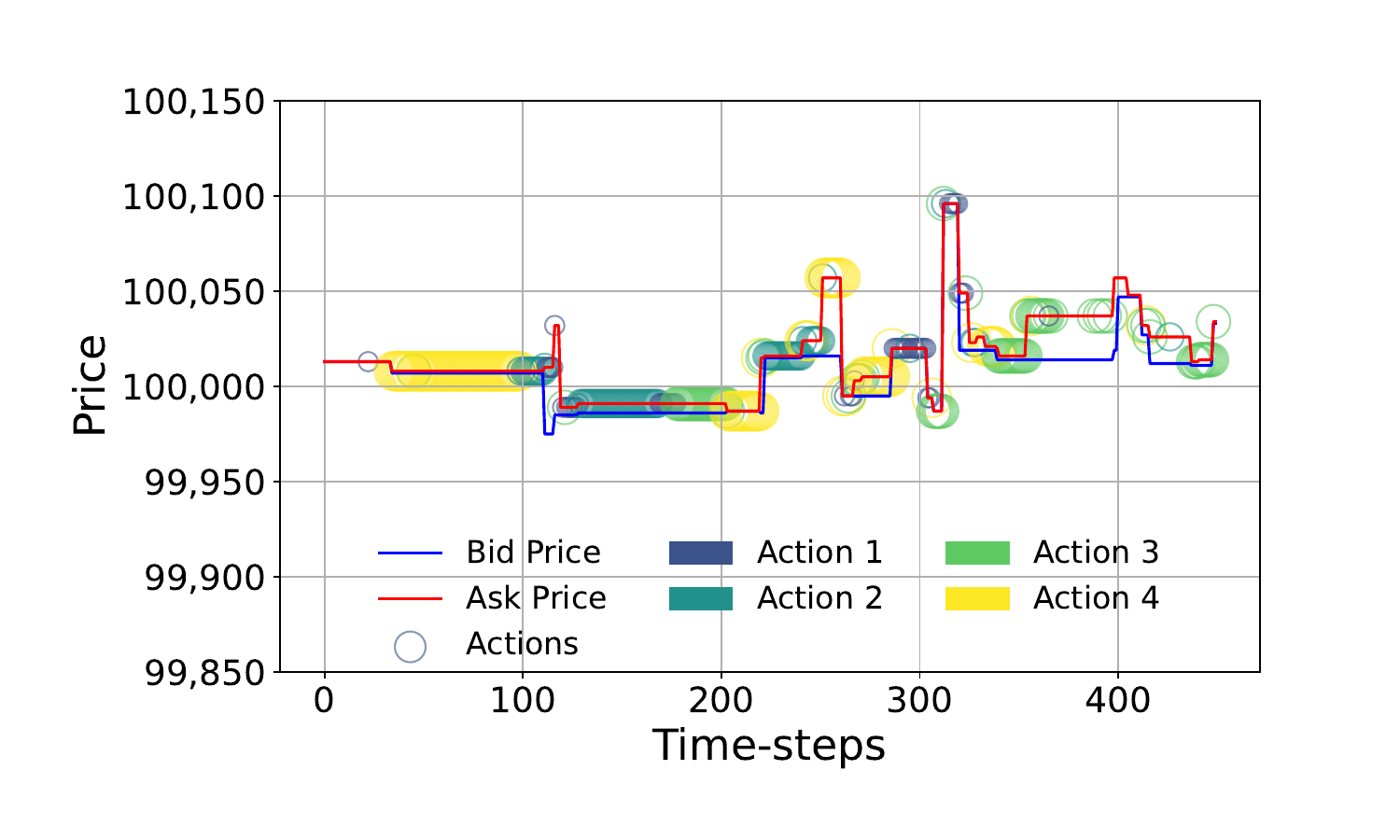}}
    \hspace{0.12cm}
    \subfloat[Execution trajectory of the RL agent: amount executed over time.\label{fig:execution_trajectory}]{\includegraphics[scale=0.23]{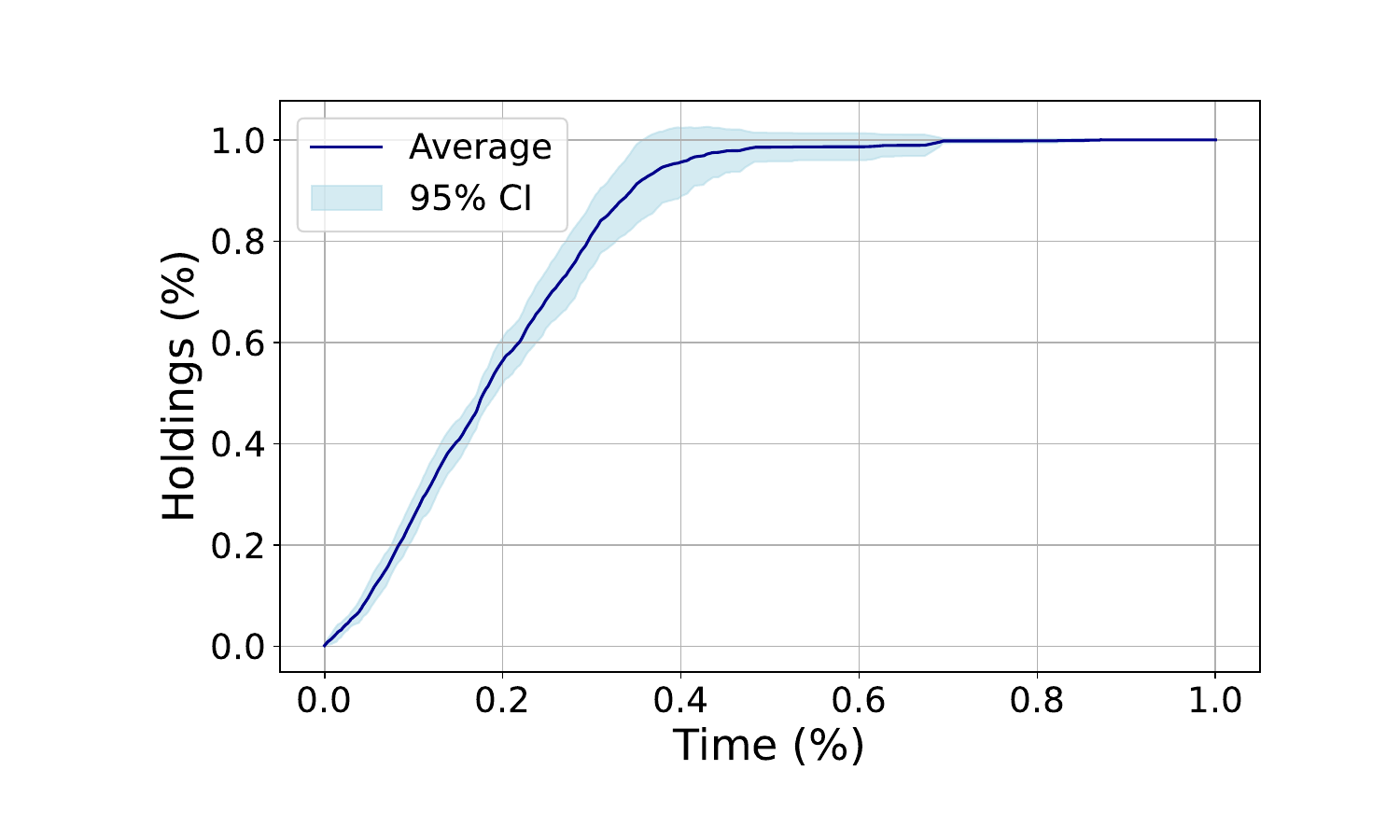}}
    \hspace{0.1cm}
    \subfloat[Temporal evolution of Q-values with respect to actions.\label{fig:q_value}]{\includegraphics[scale=0.23]{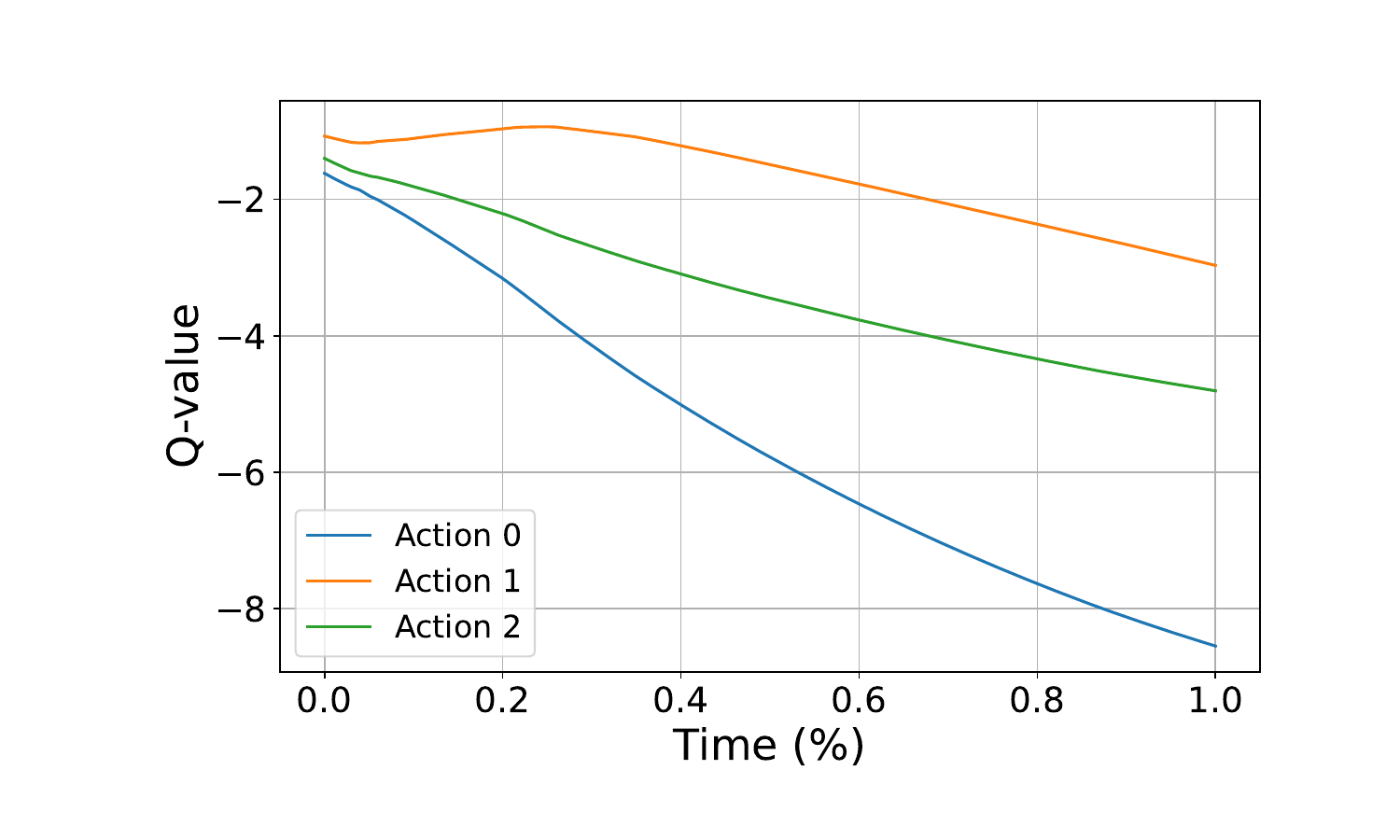}}
    \caption{RL agent behavior in a simulated execution task}
\end{figure*}

Figure \ref{fig:execution_trajectory} shows the execution trajectory of the RL agent: how much it has executed of the total size. The agent executes a significant portion of its holdings rapidly, followed by a more controlled and steady approach as the execution period progresses. This pattern aligns with optimal execution principles, balancing the trade-off between immediate market impact and long-term price stability. By front-loading the execution and then tapering off, the agent minimizes the risk of adverse market movements while efficiently managing its inventory. Figure \ref{fig:prices_actions} provides a detailed view of the bid and ask prices alongside the actions executed by the RL agent at various time steps. The figure illustrates that the actions are aligned with a minimal deviation between the bid and ask prices, indicating effective control over the execution strategy without causing market disruption.

\begin{figure}[H]
\centering
    \subfloat[Distribution of the spreads.\label{fig:spread_histogram}]{%
        \includegraphics[scale=0.24]{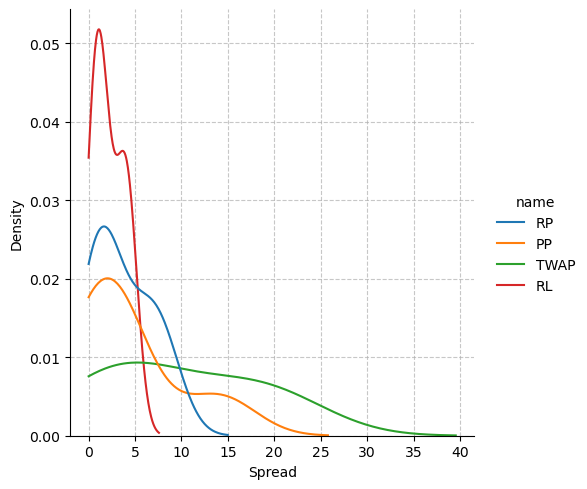}}
    \hspace{0.1cm}
    \subfloat[Distribution of the volume imbalance.\label{fig:imbalance_histogram}]{%
        \includegraphics[scale=0.24]{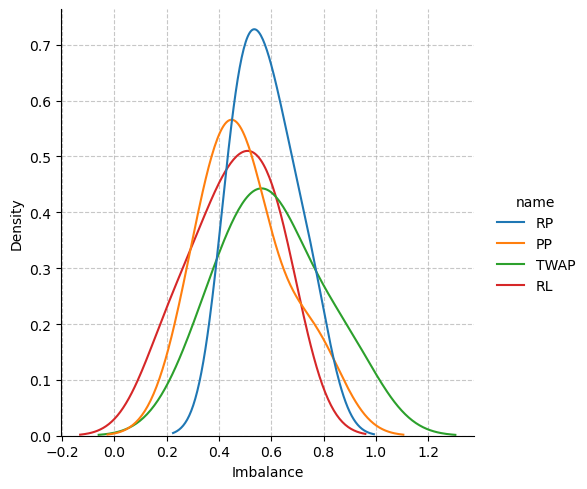}}
\caption{Empirical distributions of spreads and imbalance for different strategies.}
\label{fig:microstructure_distributions}
\end{figure}
In Figure \ref{fig:spread_histogram}, we observe that the RL strategy executes with very tight spreads, indicating that it trades at moments when liquidity is highest in the limit order book, contrarily to the other strategies. Figure \ref{fig:imbalance_histogram} shows the distribution of the volume imbalance generated by the RL agent's execution. A market impact would be visible with a skewed imabalance. In this case the imbalance is centered, indicating a consistent strategy for maintaining balance in the LOB.

\begin{table}[H]
\centering
\begin{tabular}{lcccccc}
\hline
\textbf{Strategy} & $\mathbb{E}$(IS) & $\mathbb{E}$(Pen) & $\mathbb{E}$(T) & $\sigma^2$(IS) \\ \hline
\textbf{RL}      & 0.0002  & -0.005  & 0.63 & 0.0011  \\
\textbf{TWAP}    & -0.0321  & 0.000    & 0.998 & 0.0014  \\
\textbf{Passive} & -0.0106  & -0.007  & 0.77 & 0.003  \\
\textbf{Random}  & -0.016  & -0.003   & 0.576 & 0.001  \\
\hline
\end{tabular}
\caption{Comparison of normalized performance metrics with $1000$ noise agents and $12$ momentum agents.}
\label{tab:metrics}
\end{table}
Table \ref{tab:metrics} summarizes some of the metrics also present in the plots, adding also the aggressive strategy. The first two rows represent the two parts of the reward defined in Equation \ref{eq:reward}. The last row shows on average how much time is necessary to finish the execution.

The RL agent's ability to keep the order book balanced can be attributed to its adaptive response to real-time market conditions. When encountering a growing imbalance, the agent likely adjusts its trading strategy to prevent further imbalance, thus avoiding significant deviations in asset prices. By maintaining smaller imbalances in the order book, the RL agent reduces the risk of market impact, which often leads to unfavorable price movements and increased transaction costs. 

\subsubsection{Robustness to Market Agent Configurations}
\begin{figure}[!t]
\centering
\subfloat[\scriptsize 10 noise agents\label{fig:plot1}]{\includegraphics[width=0.15\textwidth]{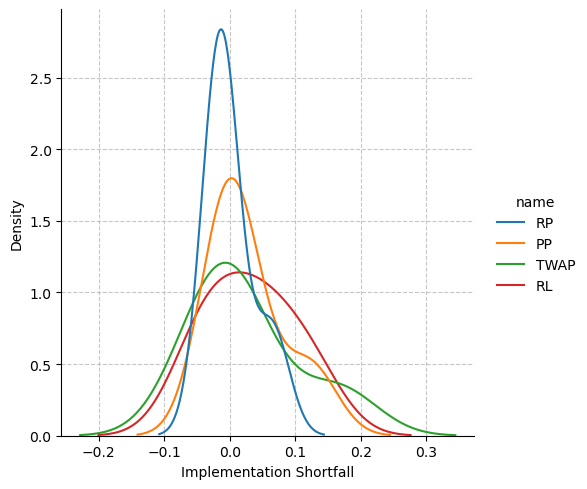}}
\hspace{0.2cm}
\subfloat[\scriptsize 1000 noise agents\label{fig:plot2}]{\includegraphics[width=0.15\textwidth]{figures/IS_1000N_12M.png}}
\hspace{0.2cm}
\subfloat[\scriptsize 2000 noise agents\label{fig:plot3}]{\includegraphics[width=0.15\textwidth]{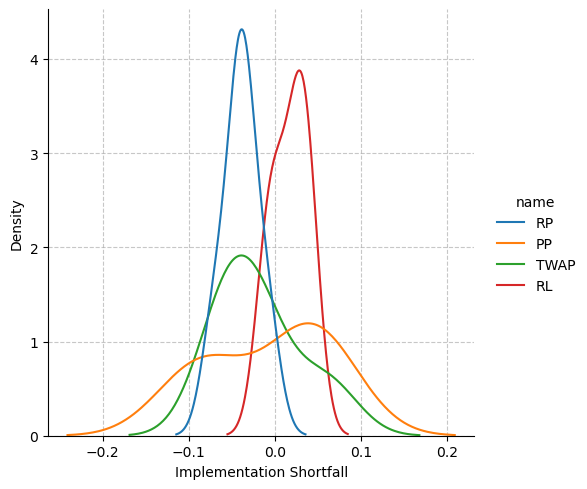}}

\vspace{0.1cm} 

\subfloat[\scriptsize 6 momentum agents\label{fig:plot7}]{\includegraphics[width=0.15\textwidth]{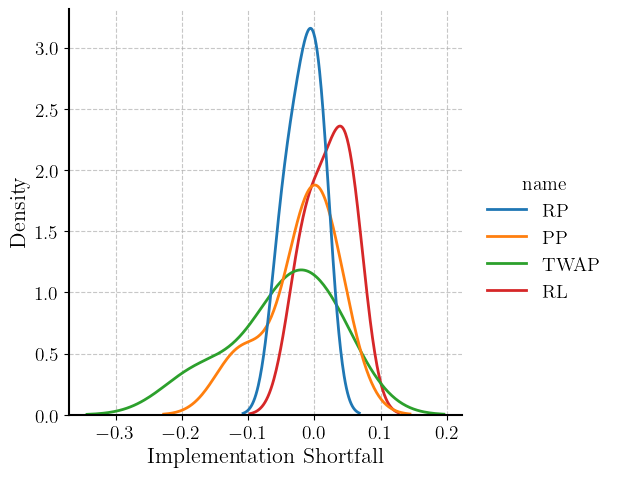}}
\hspace{0.2cm}
\subfloat[\scriptsize 12 momentum agents\label{fig:plot8}]{\includegraphics[width=0.15\textwidth]{figures/IS_1000N_12M.png}}
\hspace{0.2cm}
\subfloat[\scriptsize 24 momentum agents\label{fig:plot11}]{\includegraphics[width=0.15\textwidth]{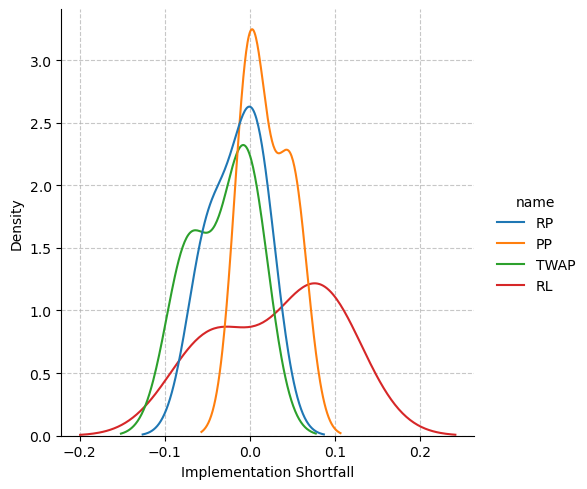}}

\caption{Implementation shortfall normalized by initial order size. Top row varies momentum agents with $12$ fixed. Bottom row varies noise agents with $1000$ fixed. }
\label{fig:12plots}
\end{figure}

Next, we demonstrate the robustness of our RL strategy by showing that it consistently outperforms all benchmark approaches in terms of both the expected implementation shortfall and the variance of performance across different market configurations. In particular, Figures~\ref{fig:12plots} and Tables~\ref{tab:metrics1} and~\ref{tab:metrics2} show the performance under varying numbers of noise agents and momentum agents.

\begin{table}[ht]
\centering
\resizebox{0.4\textwidth}{!}{%
\begin{tabular}{lccccccc}
\hline
\textbf{Strategy} & \#NA & $\mathbb{E}$(IS) & $\mathbb{E}$(Pen) & $\mathbb{E}$(T) & $\sigma^2$(IS) \\ \hline
\textbf{RL}      & 10 & -0.017  & -0.0001   & 0.55 & 0.0039  \\
                 & 2000  & 0.0173   & -0.0001 & 0.56  & 0.00043 \\
\textbf{TWAP}    & 10  & -0.067  & 0.000    & 0.998 & 0.0022  \\
                 & 2000  & -0.0184  & -0.0035  & 0.998 & 0.0024  \\
\textbf{Random}  & 10 &  -0.026 & -0.003   & 0.563 & 0.0004  \\
                 & 2000 & -0.039  & -0.002    & 0.61 & 0.00044  \\
\textbf{Passive} & 10 & -0.048  & -0.035  & 0.7 & 0.0024  \\
                 & 2000 & -0.0076  & -0.003   & 0.753 & 0.003  \\
\hline
\end{tabular}}
\caption{Comparison of normalized performance metrics with varying noise agents and $12$ momentum agents.}
\label{tab:metrics1}
\end{table}

\begin{table}[H]
\centering
\resizebox{0.4\textwidth}{!}{%
\begin{tabular}{lccccccc}
\hline
\textbf{Strategy} & \#MA & $\mathbb{E}$(IS) & $\mathbb{E}$(Pen) & $\mathbb{E}$(T) & $\sigma^2$(IS) \\ \hline
\textbf{RL}      & 6 & -0.022  & -0.001   & 0.916 & 0.0038  \\
                 & 24 & 0.004   & -0.001 & 0.51  & 0.0035  \\
\textbf{TWAP}    & 6  & -0.028  & 0.000    & 0.998 & 0.0068  \\
                 & 24  & -0.049 & -0.0015  & 0.998 & 0.024 \\
\textbf{Random}  & 6 & -0.049  & -0.002   & 0.573 &  0.0013 \\
                 & 24 & -0.036  & 0.000    & 0.54 & 0.024  \\
\textbf{Passive} & 6 & -0.024  & -0.001  & 0.79 & 0.0031  \\
                 & 24 & 0.0014  & -0.0021   & 0.71 & 0.049 \\
\hline
\end{tabular}}
\caption{Comparison of normalized performance metrics with $1000$ noise agents and varying momentum agents.}
\label{tab:metrics2}
\end{table}
We test whether RL outperforms the benchmark strategies on implementation shortfall (IS) using one-sided Student’s two-sample $t$-tests with pooled variance (per-strategy $n=50$). 
The hypotheses are 
$$H_0:\,\mathbb{E}(\text{IS})_{\text{RL}} \,\leq\, \mathbb{E}(\text{IS})_{S}
\quad \text{vs.} \quad
H_1:\,\mathbb{E}(\text{IS})_{\text{RL}} \,>\, \mathbb{E}(\text{IS})_{S},$$
where $S \in \{\text{TWAP},\;\text{Random},\;\text{Passive}\}$ denotes the benchmark strategy. Decisions are made against the one-sided 5\% critical value $t_{0.95}(98)\approx 1.660$. 

With $10$ noise agents, RL significantly outperforms TWAP ($t=4.52$) and Passive ($t=2.08$). With $2000$ noise agents, RL significantly outperforms all benchmarks: TWAP ($t=4.75$), Random ($t=8.49$), and Passive ($t=2.42$). With $6$ momentum agents, RL significantly outperforms Random ($t=2.14$).
With $24$ momentum agents, RL significantly outperforms TWAP ($t=2.26$) and Random ($t=1.81$).

Overall, RL consistently achieves statistically significant gains in the baseline and under noise-agent stress, and also demonstrates advantages in momentum-driven environments.

\section{Conclusions}\label{sec:conclusion}

The results of this study demonstrate the effectiveness of using reinforcement learning for optimal execution in financial markets. Using the DQN algorithm within a simulated market environment, we observed significant improvements in execution performance compared to the chosen benchmark strategies. Our experiments show that the RL agent consistently achieved higher returns and lower variance in implementation shortfall. The agent's ability to adapt to market conditions and execute trades close to the arrival price resulted in minimized market impact and transaction costs. The execution trajectory analysis highlighted the strategic balance of the RL agent between immediate market impact and long-term price stability. The agent managed its inventory efficiently, executing a large portion early before transitioning to a steadier pace. This behavior aligns with optimal execution principles and reflects effective inventory control.

Although our findings are promising, future work could use more tractable simulators such as zero-intelligence frameworks that capture key order book dynamics. This would provide a clearer benchmark to assess RL learning capabilities under a wider range of market conditions.

\end{document}